# Mass Generation from Higgs-like Ghosts

Scott Chapman
*Chapman University, One University Drive, Orange, CA 92866*
(Dated: November 1, 2008)

**Abstract**

Covariant quantization of gauge theories generally requires the inclusion of Fadeev Popov ghosts in the gauge-fixed Lagrangian. Normally these ghosts have fermionic statistics, but in supersymmetric theories that include fermionic gauge fields, there can also be ghosts with bosonic statistics. Since these ghosts are scalar bosons, they can have vacuum expectation values (VEVs) without violating Lorentz invariance. In particular, for the supersymmetric group SU(2/3), one can choose a gauge with a Higgs-like bosonic ghost potential that is minimized when one of the ghosts develops a VEV. This VEV breaks the SU(2) x U(1) subgroup of SU(2/3) down to U(1) and spontaneously generates vector boson masses, but without the "hierarchy" problems that accompany the Higgs field. SU(2/3) also has an SU(3) subgroup, and unitarity requires that the SU(3) gauge bosons must be confined. Since bosonic ghosts do not exist as particles asymptotically, this kind of theory could be a possible explanation if no Higgs Boson is detected at the LHC.

**Introduction**

Over the last thirty years, the Standard Model has been phenomenally successful in reproducing results from particle physics experiments and in correctly predicting the existence and properties of new particles. In fact, the only particle predicted by the Standard Model that has yet to be detected is the Higgs Boson, the scalar field needed to spontaneously generate masses for all other particles. However, in addition to a lack of direct experimental evidence, there are theoretical "hierarchy" problems associated with the "unnatural" [1] fine tuning required to renormalize the quadratic divergences of the Higgs field [2]. This unnatural fine tuning must be more precise than 1% if the Standard Model is used to describe physics at energy levels of 20 TeV or higher [3]. But even for energies as low as 1 TeV, it has been argued that the Higgs field makes the Standard Model "unnatural" [4]. It is widely believed that the Standard Model must be a low-energy approximation to a more general theory that does not have these fine-tuning "hierarchy" problems, and that the Large Hadron Collider (LHC) should uncover new physics associated with the more general theory [5].

One possible generalization of the Standard Model is Supersymmetry. In Wess-Zumino-type theories [4,6,7], the supersymmetry between bosons and fermions effectively regulates Higgs mass divergences, making these theories "natural" and applicable up to much higher energy scales [4]. Another possible generalization of the Standard Model is Technicolor [8-10]. Technicolor theories posit the existence of new interactions that create nonperturbative vacuum condensates that break the Electroweak symmetry and generate masses. These theories do not have any fundamental scalar fields (like a Higgs), so there are no "unnatural" quadratic divergences encountered during renormalization. However, Technicolor condensates are inherently nonperturbative, so they cannot be calculated directly.

This paper explores a different mechanism for mass generation that could ultimately lead to new generalizations of the Standard Model. Like Technicolor, the model considered here has no scalar fields, so there are no quadratic divergences that require unnatural fine tuning. But instead of a condensate playing the role of the Higgs to generate masses, "bosonic" ghosts play that role. These ghosts have bosonic statistics (unlike normal Fadeev Popov ghosts), and they arise when fixing the gauge of certain supersymmetric gauge theories. Due to their statistics and the fact that they have zero spin, bosonic ghosts can have a vacuum expectation value (VEV) without violating Lorentz invariance. This VEV can lead to effective masses in the same way that the Higgs VEV does in the Standard Model, but without the troublesome quadratic divergences that plague the Higgs.

The gauge theory explored in this paper is built around the supersymmetric group $SU(2/3)$. Like $SU(5)$, the fundamental representation of $SU(2/3)$ is generated by 24 Hermitian 5x5 matrices. Like $SU(5)$, 12 of $SU(2/3)$'s matrices form a $U(1) \times SU(2) \times SU(3)$ subgroup of generators that correspond to bosonic gauge fields. But unlike $SU(5)$, the remaining 12 matrices correspond to fermionic gauge fields. Throughout this paper, the 12 bosonic gauge directions will be referred to as being "even" and the 12 fermionic directions as "odd". These labels are motivated by the commutation relations presented below.

SU(2/3) is similar to the group SU(2/1), which was first presented in the literature in 1979 in an attempt to explain why the Weinberg angle was so close to 30 degrees. In [11,12] it was shown that if one added 2 anti-commuting Grassman dimensions to the usual four dimensions of spacetime, then the "odd" gauge fields with Grassman indices would be bosonic Lorentz scalars with regard to the usual four spacetime indices. As a result, they could behave like a Higgs field, developing a vacuum expectation value (VEV) and generating masses for the vector bosons. Later works pointed out a number of problems with this theory, including violations of the spin-statistics theorem and the fact that the "U(1)" part of the Lagrangian had the "wrong" sign [13]. This wrong sign made it impossible to mix the U(1) field with the diagonal SU(2) field in the usual way, and it also implied that the U(1) field either had negative kinetic energy or an indefinite metric that would prohibit the existence of asymptotic photons. Later, it was also pointed out that the "Weinberg angle" of the theory was actually 45 degrees [14], not 30 degrees as previously thought. Eventually SU(2/1) was mostly abandoned as a potential theory for the ElectroWeak interaction, although there is still some work being done in that area [15].

Recently, the negative kinetic energy of the U(1) field has led to the reintroduction into the literature of SU(2/1) as a possible theory for cosmological phantom energy [16]. This more recent work assumes only the four normal spacetime dimensions (no Grassman dimensions). It also formulates the theory with the odd (fermionic) gauge fields being spin-1 vectors. Problems with violation of the spin-statistics theorem are avoided by assuming that the fermionic, spin-1, odd fields are always bound in bosonic doublets.

The formulation of SU(2/3) in this paper also involves just the normal four spacetime dimensions and vector fields for both the even and odd gauge directions. Fixing the gauge of the theory produces the usual fermionic ghosts associated with the bosonic gauge fields, but it also produces bosonic ghosts associated with the fermionic gauge fields. Like the bosonic fields of [11,12,15], these bosonic ghosts have the right spin and statistics to behave like Higgs bosons, but they do not require additional Grassman spacetime dimensions. Using the Antifield formalism [17,18], it is shown that one can pick a gauge with a Higgs-like potential for the bosonic ghosts that is minimized when they develop a VEV. This VEV produces vector boson masses and mixing that imply a "Weinberg angle" of 30 degrees.

**The Group SU(2/3)**

The group SU(2/3) is generated by 24 5x5 matrices $\lambda_A$, where the index $A$ runs from 0 to 23. For indices 1-23, the matrices are the same as SU(5) matrices, and they are grouped into blocks as follows:

$$\left( \begin{array}{c|c} \lambda_i \in SU(2) & \lambda_\beta \\ \hline \lambda_\beta & \lambda_a \in SU(3) \end{array} \right) \qquad \begin{array}{l} i,j,k \in \{1-3\} \\ \alpha,\beta,\gamma \in \{4-15\} \\ a,b,c \in \{16-23\} \end{array} \qquad (1)$$

where the index conventions used in this paper are also shown. In addition to those 23 matrices, there is also a diagonal U(1) matrix,

$$\lambda_0 = \sqrt{3}\,\mathrm{diag}\left(\tfrac{1}{2},\tfrac{1}{2},\tfrac{1}{3},\tfrac{1}{3},\tfrac{1}{3}\right). \qquad (2)$$

All of the generators have vanishing supertraces, where the supertrace for this theory is defined by adding the first two diagonal entries of a matrix and subtracting the last three. The indices $\alpha,\beta,\gamma \in \{4-15\}$ are labeled as "odd", and all other indices are "even". Just as in SU(5), the "even" generators close in a U(1) x SU(2) x SU(3) subgroup. An SU(2/3) gauge transformation takes the form $U = \exp(i\lambda_A \varepsilon_A)$, where $\varepsilon_A$ is a commuting variable for "even" indices and an anticommuting (Grassman) variable for "odd" indices.

Since SU(2/3) is a closed group, one can define structure constants $f^C_{AB}$ through the following equations:

$$\{\lambda_A, \lambda_B\} = i f^C_{AB} \lambda_C \qquad \text{for } A \text{ \underline{and} } B \in \{4-15\} \text{ (both are ``odd'' indices)}$$

$$[\lambda_A, \lambda_B] = i f^C_{AB} \lambda_C \qquad \text{for } A \text{ \underline{or} } B \notin \{4-15\} \text{ (at least one index is ``even'')}. \qquad (3)$$

The top relation shows that the anticommutator of two "odd" fields makes an "even" field. This, combined with the fact that the "even" generators close in their own subgroup, is the motivation for the "even" and "odd" naming convention. The covariant derivative and field strength tensor are given by:

$$D_\mu \equiv \partial_\mu - ig A^A_\mu \lambda_A$$

$$g\lambda_C F^C_{\mu\nu} \equiv i[D_\mu, D_\nu]$$

$$= g\lambda_C \left( \partial_\mu A^C_\nu - \partial_\nu A^C_\mu + g f^C_{AB} A^A_\mu A^B_\nu \right). \qquad (4)$$

The gauge invariant Lagrangian for the theory is given by a supertrace:

$$\mathcal{L}_0 = -\tfrac{1}{2}\mathrm{str}\!\left(\lambda_A F^A_{\mu\nu}\lambda_B F^{B\mu\nu}\right)$$
$$= -\tfrac{1}{4}F^0_{\mu\nu}F^{0\mu\nu} - \tfrac{1}{4}F^i_{\mu\nu}F^{i\mu\nu} + \tfrac{1}{4}F^a_{\mu\nu}F^{a\mu\nu} + \tfrac{\sqrt{3}}{2}if^\alpha_{0\beta}F^\alpha_{\mu\nu}F^{\beta\mu\nu}. \tag{5}$$

Here one can see that the SU(3) part of the Lagrangian (the third term above) has the "wrong" sign, leading either to negative kinetic energy or to the existence of negative norm states for those fields. This is analogous to the SU(2/1) case in which the U(1) fields have the wrong sign. In SU(2/3), unlike SU(2/1), since both the U(1) and SU(2) fields have the "right" sign, it is possible for gauge boson mass generation to result in mixing between the U(1) and SU(2) gauge fields. This will be shown explicitly below.

As usual, to quantize the theory a gauge must first be chosen. In the path integral approach, this is accomplished by adding ghost terms to the Lagrangian then functionally integrating over both gauge fields and ghosts in order to calculate observables. One of the most direct ways to determine the allowed Lorentz-invariant ghost terms is to employ the Antifield formalism [17,18]. In this formalism, the Hilbert space is enlarged to include an antifield for every field, and the general form of the gauge-fixed action is determined by solving a cohomological "master equation". To specify a particular gauge, one chooses a "gauge-fixing fermion", then redefines the antifields in terms of its derivatives.

For SU(2/3), the solution to the Antifield master equation adds the following terms to the Lagrangian:
$$\mathcal{L}^* = A^{*\mu}_A D^A_{\mu B} C^B + \tfrac{1}{2} g f^A_{BC} C^B C^*_A C^C - g f^A_{\beta\gamma} C^\beta C^*_A C^\gamma + i\overline{C}^*_A b^A$$
$$D^A_{\mu B} \equiv \partial_\mu \delta^A_B + g f^A_{CB} A^C_\mu \tag{6}$$

The $C^A$ fields are ghosts that have statistics opposite to their corresponding gauge fields. In other words the even ghosts have fermionic statistics, but the odd ghosts $C^\alpha$ have bosonic statistics. The starred fields are "antifields" that can be rewritten in terms of normal fields by using a gauge-fixing fermion $\psi$ and the equations

$$A^{*\mu}_A = \frac{\delta^R \psi}{\delta A^A_\mu}, \qquad C^*_A = \frac{\delta^R \psi}{\delta C^A}, \qquad \overline{C}^*_A = \frac{\delta^R \psi}{\delta \overline{C}^A}, \tag{7}$$

where derivatives are taken from the right.

The gauge chosen in this paper is defined through the following gauge-fixing fermion:

$$\psi = \int d^4x \left\{ 2i \, \text{str}\left[\lambda_A \overline{C}^A \lambda_B \left(\partial^\mu A_\mu^B - \tfrac{1}{2}\xi b^B\right)\right] \right.$$

$$\left. + \tfrac{1}{2} A_\mu^\alpha \left(D_\beta^{\mu\alpha} - 2gf_{b\beta}^\alpha A^{\mu b}\right) C^\beta + \frac{i\sqrt{3}}{g} C^0 \left(\lambda(C^\alpha)^2 - 2m^2\right)\right\}. \tag{8}$$

Using the first term of (8) in (6) and (7), then integrating out the $b^A$ fields, one obtains the usual Lorentz gauge fixing and ghost terms. But the second line of (8) produces new terms, including a Higgs-like contribution to the gauge-fixed Lagrangian:

$$\mathcal{L}_H = \tfrac{1}{2}\left(D_\beta^{\mu\alpha} - 2gf_{b\beta}^\alpha A^{\mu b}\right) C^\beta D_{\mu\gamma}^\alpha C^\gamma + \tfrac{1}{2}m^2 (C^\alpha)^2 - \tfrac{1}{4}\lambda((C^\alpha)^2)^2. \tag{9}$$

The presence of these terms is quite remarkable. One starts with a pure gauge theory involving no scalar fields, and just by fixing the gauge, one obtains a Higgs-like contribution involving scalar bosonic ghost fields.

In order to minimize the Higgs potential, at least one component of the $C^\alpha$ field should develop a vacuum expectation value (VEV). We shall assume that only $C^{12}$ develops a VEV:

$$\langle C^{12} \rangle = v = \frac{m}{\sqrt{\lambda}}, \quad \text{where} \quad \lambda_{12} \equiv \frac{1}{2}\begin{pmatrix} 0 & 0 & 0 & 0 & 1 \\ 0 & 0 & 0 & 0 & 0 \\ 0 & 0 & 0 & 0 & 0 \\ 0 & 0 & 0 & 0 & 0 \\ 1 & 0 & 0 & 0 & 0 \end{pmatrix} \tag{10}$$

Due to this VEV, $\mathcal{L}_H$ gives effective masses to some of the vector bosons. In particular, one obtains the following effective mass terms for the SU(2) x U(1) fields:

$$\mathcal{L}_M = \tfrac{1}{2} M_W^2 \left[(A_\mu^1)^2 + (A_\mu^2)^2\right] + \tfrac{1}{2} M_Z^2 (\sin\theta \, A_\mu^0 + \cos\theta \, A_\mu^3)^2, \tag{11}$$

with

$$M_W \equiv \tfrac{1}{2} gv \qquad M_Z \equiv \frac{M_W}{\cos\theta} \qquad \sin^2\theta = \tfrac{1}{4}. \tag{12}$$

Equation (11) has a similar form to that of the mass terms in the standard Electroweak model, with a "Weinberg angle" equal to 30 degrees.

Although the vector boson mass ratios look promising, SU(2/3) does not appear to be a good candidate for an SU(5)-like generalization of the Standard Model. A major problem is that the "photon" in this model corresponds to the following SU(2/3) generator:

$$\lambda_{ph} = \text{diag}\left(\tfrac{1}{2}, 1, \tfrac{1}{2}, \tfrac{1}{2}, \tfrac{1}{2}\right). \tag{13}$$

Since this generator does not have any zeros on the diagonal, it implies that any fermions in the fundamental **5** representation of SU(2/3) must have some "electric" charge. In other words, unlike in SU(5) theories, there cannot be a neutrino in the fundamental representation. Other problems with SU(2/3) include wrong charges for the quarks and the fact that some of the SU(3) fields also get masses, breaking that symmetry. Nonetheless, SU(2/3) is an instructive model to see how bosonic ghosts can generate masses for vector bosons.

A general feature of any supersymmetric theory involving bosonic ghosts is that part of the Lagrangian will have the wrong sign. In papers on SU(2/1), the U(1) field is the one with the wrong sign, and this is usually interpreted to mean that those fields (or in the present case, the SU(3) fields) have negative kinetic energy [13,16]. Theories with negative kinetic energy, however, do not have a stable vacuum unless there is some kind of cutoff, and that cutoff cannot be Lorentz-invariant [19]. The only other option for a theory with the "wrong" sign on its kinetic terms is to choose an indefinite-metric state representation, and that is the option chosen here. In that approach, the eigenvalues of $A_\mu^a$ become imaginary [20], so one can make the replacement $A_\mu^a \to iA_\mu^a$ in the Lagrangian, and the kinetic energy becomes positive. But the indefinite metric introduces states with negative norm, so in order to preserve unitarity, one has to restrict physical states to only those with positive norms.

In the context of plane wave expansions, the restriction to positive-norm states translates into the restriction that physical states can only contain an even number of SU(3) gauge bosons (called "gluons" from now on for convenience). This means, for example, that no single "gluon" can exist as an asymptotically free state. Fortunately, if one starts with this severe restriction on physical states, then in the limit of vanishing coupling, the Hamiltonian does not create or destroy an odd number of gluons, so the restriction is consistent. If one now turns up the coupling, single gluons will be created and destroyed, and negative norm components will be mixed into the positive norm states. Adjusting the norms of these states back to +1 implicitly imparts more weight to interacting states and effectively strengthens the coupling constant. As more negative-norm states are mixed in, the effective coupling explodes before the overall norm of a state would become negative. In other words, an indefinite metric and the compulsory restriction to positive norm states leads to gluon confinement and the SU(3) coupling being effectively stronger than the other coupling constants.

Supersymmetric theories of the type considered here generally have another quirk: some of the fields have integral spin but fermionic statistics. This spin-statistics problem was addressed in [16] by proposing that physical states would only consist of fermionic gauge fields bound into pairs, never alone or in odd-number combinations. These bound states would then have the right relationship between spin and statistics, and it was proposed that they could be a candidate for dark matter. Since "gluons" cannot exist asymptotically and the fermionic gauge fields are "colored", these bound states may be restricted to exist in combinations that would only interact weakly with other matter or forces.

Even though SU(2/3) has the wrong "photon" to be a successful generalization of the Standard Model, it is possible that other supersymmetric groups could be more successful in this regard. For example, consider the group SU(2/4). Following the same approach as used for SU(2/3), one can choose a gauge that breaks the U(1) x SU(2) x SU(4) subgroup down to U(1) x SU(3). All of the original SU(4) gauge bosons would be confined (but just the SU(3) subgroup of them would remain massless). If a variation of gauge choice can be found that results in a massless photon corresponding to the generator $\lambda_{ph} = \mathrm{diag}\left(0,1,\frac{1}{3},\frac{1}{3},\frac{1}{3},0\right)$, then one would be able to fit a full family of 15 quarks and leptons (excluding the right-handed neutrino) into a single anti-symmetric representation.

A possible signature of this kind of supersymmetric theory would be the non-detection of a Higgs Boson at the LHC. In the Standard Model and the usual Wess-Zumino-type supersymmetric extensions to it, Higgs Bosons can and should exist as asymptotic particle states detectable at the LHC. But in the kind of supersymmetric gauge theory considered here, bosonic ghosts most likely cannot exist as asymptotic particle states. This should be verified theoretically by canonically quantizing a supersymmetric gauge theory using the BRST approach [17,21]. That approach requires physical states to be annihilated by a BRST operator that commutes with the Hamiltonian. In standard gauge theories, this BRST constraint in the context of Lorentz-invariant gauges leads to the restriction that ghosts cannot exist as asymptotic particles.

Another advantage of studying BRST quantization is that it can provide a mechanism to generate masses for leptons and quarks, which have not been included in the present analysis.

Since the BRST operator contains leptons and quarks along with ghosts, a judicious choice of gauge can introduce bosonic ghost interactions with the matter fields. When the bosonic ghost develops a VEV, this interaction can lead to quark and lepton masses.

**Summary**

Mass generation in the Standard Model has severe theoretical "hierarchy" problems caused by renormalization of the Higgs Boson mass. One of the main motivations of many extensions to the Standard Model has been to avoid these problems. This paper suggests a possible new way to avoid hierarchy problems by assuming the existence of a supersymmetric theory that includes bosonic ghosts, then using those ghosts to break symmetries and generate masses. An intriguing consequence of these theories is that they must include some gauge bosons that are confined and experience an effective coupling that is stronger than that of the unconfined fields. In addition, these theories predict bound states of fermionic gauge fields that could be good candidates for dark matter. A possible experimental signature for this kind of mass-generation mechanism would be the non-detection of a Higgs Boson at the LHC.

**References**
[1] G. 't Hooft, *Recent developments in gauge theories (NATO ASI Series B: Physics Vol. 59)*, pages 135-157, Plenum Press (1979).
[2] E. Gildener, *Phys. Rev.* **D14**, 1667 (1976).
[3] C. Kolda and H. Murayama, *JHEP* **0007**, 35 (2000).
[4] P. Binétruy, *Supersymmetry: Theory, Experiment, and Cosmology*, Oxford University Press (2006).
[5] J. Ellis, http://cerncourier.com/cws/article/cern/29893; J. Ellis. (http://arxiv.org/abs/hep-ph/0611237).
[6] J. Wess and B. Zumino, *Nucl. Phys.* **B70**, 39 (1974).
[7] R. Haag, J. Lopuszanski, and M. Sohnius, *Nucl. Phys.* **B88**, 257 (1975).
[8] S. Weinberg, *Phys. Rev.* **D19**, 1277 (1979).
[9] L. Susskind, *Phys. Rev.* **D20**, 2619 (1979).
[10] C.T. Hill and E.H. Simmons, *Phys. Rept.* **381**, 235 (2003).
[11] Y. Ne'eman, *Phys. Lett.* **B81**, 190 (1979).
[12] P.H. Dondi and P.D. Jarvis, *Phys. Lett.* **B84**, 75 (1979).
[13] R.E. Eccelstone, *J. Phys.* **A13**, 1395 (1980).
[14] R.E. Eccelstone, *Phys. Lett.* **B116**, 21 (1982).
[15] J. Thierry-Mieg. arXiv:0806.4628 (2008).
[16] M. Chaves and D. Singleton, *Mod.Phys.Lett.* **A22**, 29-40 (2007).

[17]   M. Henneaux and C. Teitelboim, *Quantization of Gauge Systems,* Pinceton University Press (1992).
[18]   A. Fuster, M. Henneaux, and A. Maas, *Int.J.Geom.Meth.Mod.Phys.* **2**, 939-964 (2005).
[19]   J. Cline, S. Jeon, and G. Moore, *Phys.Rev.* **D70**, 043543 (2004).
[20]   W. Pauli, *Rev. Mod. Phys.* **15**, 175 (1943).
[21]   R. Marnelius, *Nucl. Phys.* **B395**, 647 (1993).